\documentclass[conference,a4paper]{IEEEtran}
\usepackage[pdftex]{graphicx}
\usepackage{amsmath,amssymb}

\def\(({\left(}
\def\)){\right)}
\def\[[{\left[}
\def\]]{\right]}

\newcommand{\be}{\begin{equation}}
\newcommand{\ee}{\end{equation}}
\newcommand{\bea}{\begin{eqnarray}}
\newcommand{\eea}{\end{eqnarray}}

\let\a=\alpha

 \let\D=\Delta   
         
 \let\ee=\epsilon \let\r=\rho 

\def\ie{{i.e. }}

\def\to{\rightarrow}

\newcommand{\beq}{\begin{equation}}
\newcommand{\eeq}{\end{equation}}

\begin{document}

\sloppy

\title{Properties of spatial coupling in compressed sensing}

\author{
 \IEEEauthorblockN{Francesco Caltagirone}  \IEEEauthorblockA{Institut de Physique Th\'eorique\\
    CEA Saclay and URA 2306, CNRS\\ 91191 Gif-sur-Yvette,
    France.}
\and \IEEEauthorblockN{Lenka
    Zdeborov\'a}  \IEEEauthorblockA{Institut de Physique Th\'eorique\\
    CEA Saclay and URA 2306, CNRS\\ 91191 Gif-sur-Yvette,
    France.}}


\maketitle

\begin{abstract}
In this paper we address a series of open questions about the construction of spatially coupled measurement 
matrices in compressed sensing. For hardware implementations one is
forced to depart from the limiting regime of parameters in which the
proofs of the so-called threshold saturation work. We investigate
quantitatively the behavior under finite coupling range, the
dependence on the shape of the coupling interaction, and optimization
of the so-called seed to minimize distance from optimality. Our
analysis explains some of the properties observed empirically in
previous works and provides
new insight on spatially coupled compressed sensing.
\end{abstract}

\section{Introduction}

Spatial coupling is a methodological concept that permits to enhance
considerably limits of algorithmic tractability in a wide range of inference
problem where the underlying graphical model can be designed. Spatial
coupling was originally introduced in low density parity check (LDPC) codes
by~\cite{FelstromZigangirov99}, followed by a line of theoretical work
on (terminated) convolutional LDPC codes,
notably \cite{lentmaier2005terminated,lentmaier2010iterative}.  
The term itself, as well as a large part of the theoretical proofs and understanding comes
from the work of \cite{KudekarRichardson10}. The fact that with spatial coupling design of the system of interest one obtains
tractable algorithms working down to the corresponding information
theoretical limits (a phenomenon termed ``threshold saturation'') has been
successfully applied in a number of problems in error correcting codes, signal
processing, communication systems and computer science problems, for
an incomplete list of examples see references in
\cite{KudekarRichardson12}. 

One particularly important problem where spatial coupling was
successfully applied is compressed sensing. It is well known that most signals of interest are compressible and
this fact is widely used to save storage place. But usually compression is made only once
the full signal is measured. In many applications (e.g. medical imaging using MRI or
tomography) it is desirable to reduce the measurement time as much as possible (to
reduce costs, radiation exposition dose etc.). Hence it is desirable to measure the signal
directly in the compressed format. Compressed sensing is a concept implementing
exactly this idea by designing new measurement protocols and sparse signal
reconstruction algorithms \cite{Donoho:06,Candes:2008}.

The first investigation of spatial coupling in compressed sensing is
due to \cite{KudekarPfister10} where some limited improvement of
performance was observed. A combination of Bayesian probabilistic
approach, the use of approximate message passing algorithm and a
special design of the (spatially coupled and seeded) measurement matrix led to an empirical evidence
and a conjecture that with
spatial coupling the information theoretic limits can be indeed achieved in
compressed sensing \cite{KrzakalaPRX2012,KrzakalaMezard12}. This
conjecture was proven rigorously by \cite{DonohoJavanmard11}. Recent
generic and simple proof of the so-called threshold saturation also
applies to the case of compressed sensing \cite{YedlaJian12}. 

The view of possible experimental implementations of spatially coupled
measurement matrices in compressed sensing devices (e.g. optical
imagers using highly flexible arrays of micro-mirrors) motivates the need
of more detailed study of this subject. 
The main open question that needs to be resolved concerns the optimization of the
spatially coupled measurement matrix for systems of finite size. Such
a process includes optimization of the number of blocks, range
and shape of interactions. It also includes minimization of the ``termination cost'',
i.e. the determination of the size and strength of the seed that minimizes the final undersampling
rate while assuring a robust propagation of the ``successful
reconstruction'' wave. In this paper we address some of these questions.



\section{Spatial coupling for compressed sensing}

In compressed sensing we consider a $N$-dimensional signal (vector)
${\bf x}$, this vector is $K$-sparse, i.e. only $K<N$ of its elements
are nonzero. The measurements device records $M<N$ linear projections of
this signal into a $M$-dimensional vector ${\bf y}$. These projections
can be represented by a $M\times N$ measurement matrix $F$, such that
${\bf y} = F {\bf x} + {\bf \xi}$ where ${\bf \xi}$ is (element-wise) an additive
white Gaussian noise of variance~$\Delta$. The goal is to reconstruct tractably
the sparse vector ${\bf x}$ from the knowledge of ${\bf y}$ and $F$
for a given sparsity $\rho = K/N$ and the lowest possible undersampling ratio $\alpha = M/N$.  
Here we consider a model case of compressed sensing where the non-zero
elements of the signal ${\bf x}$ are taken as independent normally
distributed variables of zero mean and unit variance, and we
consider the asymptotic limit $N\to \infty$. We aim to minimize the mean
squared error between the reconstructed and true signal
$E=\sum_{i=1}^N (x^{\rm inferred}_i - x^{\rm true}_i)^2/N$.

To reconstruct the signal we work in the Bayesian setting where the goal is to sample the
posterior probability distribution $P({\bf x}| F, {\bf y}) = P({\bf
  x}) P({\bf y}| F, {\bf x}) /Z$. Such sampling is in general
computationally difficult ($\# P$-complete). An influential line of
work \cite{DonohoMaleki09,Rangan10b} introduced the approximate message passing (AMP), an iterative
algorithm closely related to belief propagation (BP), as an 
algorithm that performs this sampling efficiently as long as the
elements of the matrix $F$ are random independent variables and the
system size $N$ is large. Similarly to BP decoding in
LDPC codes the AMP only works for some region of parameters $\rho$,
$\alpha$, $\Delta$. In the regime of low measurement noise $\Delta$ there is an algorithmic threshold $\alpha_{\rm
  BP}(\Delta,\rho)$ below which the algorithm fails, yet a good reconstruction of the
signal is in principle possible down to an undersampling rate
$\alpha_c(\Delta,\rho)< \alpha_{\rm BP}(\Delta,\rho)$. The values of
$\alpha_{\rm BP}$ were obtained rigorously via state evolution
asymptotic analysis of the AMP algorithm
\cite{bayati2012universality}, the values $\alpha_c$ are known
non-rigorously thanks to the replica calculation
\cite{KrzakalaPRX2012,KrzakalaMezard12}. 

The remarkable result of \cite{KrzakalaPRX2012,KrzakalaMezard12,DonohoJavanmard11} is that when
the concept of spatial coupling is used to design the measurement matrix $F$ then the
AMP algorithm reconstructs successfully down to measurement ratios
$\alpha_c$ (this is analogous to the ``threshold saturation'' in LDPC
codes). 

\begin{figure}
\centering
\includegraphics[scale = 0.2]{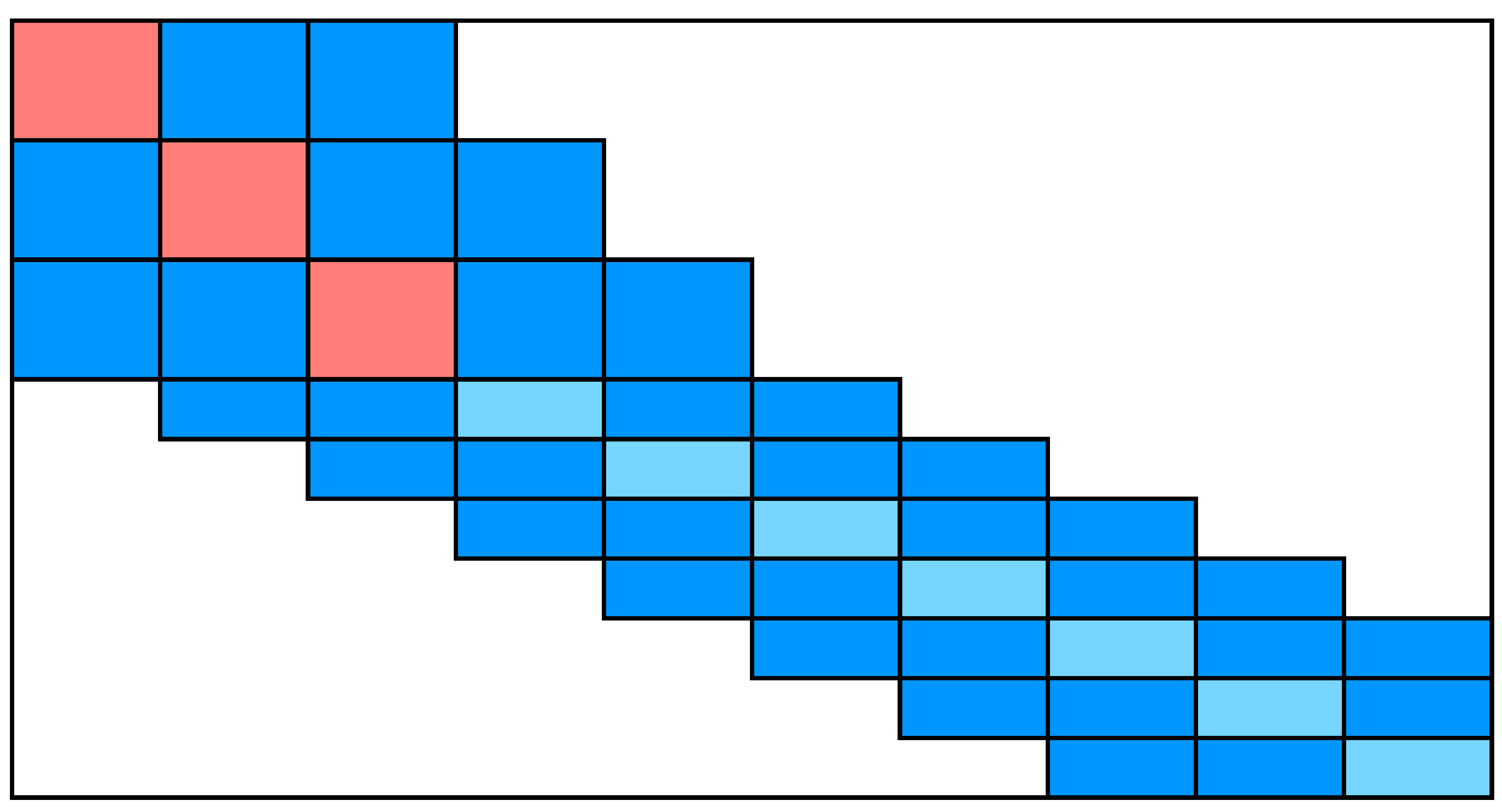}
\caption{A schematic representation of the seeded spatially coupled
  measurement matrix $F$. In the example 
above the number of blocks is $L=10$, the interaction range is $w=2$, the seed size is 
$w_s=3$, the bulk and seed undersampling ratios are, respectively, $\a_b=0.4$ and $\a_s=0.8$. 
In light blue are the bulk blocks, in red the seeding blocks, in blue the interaction blocks, and in white the null elements.}
\label{matrix}
\end{figure}

The seeded spatially coupled matrix $F$ used in this work is defined as follows, see
Fig.~\ref{matrix}. We split the $N$ components of the signal ${\bf x}$ into $L$
equally sized blocks. We then split the components of the
measurements ${\bf y}$ into $L$ blocks, each of the first $w_s$ of them will
have size $\alpha_s N/L$ (we call these first $w_s$ blocks the ``seed''), each of the remaining blocks will have
size $\alpha_b N/L$ (we call these remaining blocks the ``bulk''). We choose the elements of each of the block to be
normally distributed with zero mean and variance $J_{qr}/N$, where
$q,r=1,\dots,L$.  The
coupling matrix $J_{qr}$ is taken as 
\begin{align}
&J_{qr}=\frac{J}{w}g\left( \frac{r-q}{w} \right)\, , \quad {\rm
  where}\quad   \frac{1}{w} \sum_{z=-w}^wg\left(\frac{z}{w}\right)=1\, ,
\nonumber \\
&g\left(\frac{z}{w}\right)=0 \, , \quad {\rm for} \quad  z\in
(-\infty,-w) \cup (w,\infty)\, .
\label{interact}
\end{align}

This definition is very similar though not identical to the
definitions in the previous literature on spatial coupling in
compressed sensing. Note that \cite{KrzakalaPRX2012} treated only the
case of $w_s=w=1$, in \cite{KrzakalaMezard12} generic $w$ was
considered and other designs of the matrix suggested, but without a
systematic study of their advantages. The theoretical proofs of
threshold saturation \cite{DonohoJavanmard11,YedlaJian12} deal with
the limit $1 \ll w \ll L \ll N$ and the seed is either not considered
at all or has far too big cost for a practical implementation. In this
limit the reconstructed part of the signal propagates as a kind of a
travelling wave into the whole system \cite{kudekar2012wave}. 
Given that the system sizes of
practical interest range from about $N=10^3$ to say $N=10^7$
(depending on the application) this limit
cannot be implemented very closely. The main goal of this work is to
still consider infinite block size, and to study the dependence on the
values of parameters $w$, $L$, $w_s$, $\alpha_s$ and the function $g(\cdot)$ in order to achieve
reconstruction of the signal under smallest
possible average undersampling ratio
\beq
\a_{\mathrm{eff}}=\frac{\a_b (L-w_s)+\a_s w_s}{L} \, .
\eeq

When the size of every block is infinite the state evolution equations
were derived in \cite{KrzakalaPRX2012,DonohoJavanmard11} to describe
the asymptotic evolution of the mean-squared error $E_p$ in each
block
 \beq
\begin{split}
&\hat{q}^{t+1}_p=  \sum_{q=1}^{L} \frac{\a_{q} J_{qp}}{\Delta +\sum_{r=0}^{L} J_{q r} E^{t}_r}\ , \\
&E_p^{t+1} =\rho-\frac{\rho^2 \hat{q}^{t+1}_p}{\hat{q}^{t+1}_p +1}
G\left( \r,\hat{q}^{t+1}_p\right)\, ,
\end{split}
\eeq
where
\beq
G\left( \r,\hat{q}\right)=\int \frac{dz}{\sqrt{2\pi}} \frac{ z^2\,
  e^{-\frac{z^2}{2}}}{\rho + (1-\rho)e^{-\frac{z^2\hat{q}}{2}}
  \sqrt{\hat{q} +1} }\, .
\eeq
We hence study the behavior of these state evolution equations. The
only initialization that is algorithmically possible in practice is
$E_p=\rho$ (or larger) for every block $p$. This is the main reason
why we need the seed to exist - its higher undersampling ratio
$\alpha_s$ has to ensure that at least the first $w_s$ blocks will get
reconstructed and the interaction of range $w$ then has to ensure this
reconstruction will propagate into the whole system.  

\section{The finite-$w$ phase transition} 

In the limit of many blocks $L\to \infty$, given that the seed remains
of finite size, one has that the 
effective measurement ratio is equal to the bulk one, $\lim_{L\to \infty}\a_{\mathrm{eff}}=\a_b$. This, combined with the limit 
$w\to \infty$ taken after $L\to \infty$ is used in the proofs of
threshold saturation, i.e. optimal recovery of the 
signal is achieved down to $\a_c$. In practical implementations, the
interaction range $w$ has to be relatively small since $w < L < N$,
and also as we will see since the seed cost grows with $w$. 
In this Section we show, by analyzing numerically the state evolution equations, that if $w$ is finite reconstruction is 
possible only down to some $\a_w > \a_c$ {\it even} when $L\to
\infty$. The seed we use for this analysis is very large and very
strong $w_s\gg 1$, $\alpha_s>1$, the number of blocks $L$ is large, and a flat
interaction $g(x)=1/2$ is used. The results then depend only on
the parameters $\alpha_b$, $\rho$, $\Delta$, and $w$.

\begin{figure}
\centering
\includegraphics[scale = 0.7]{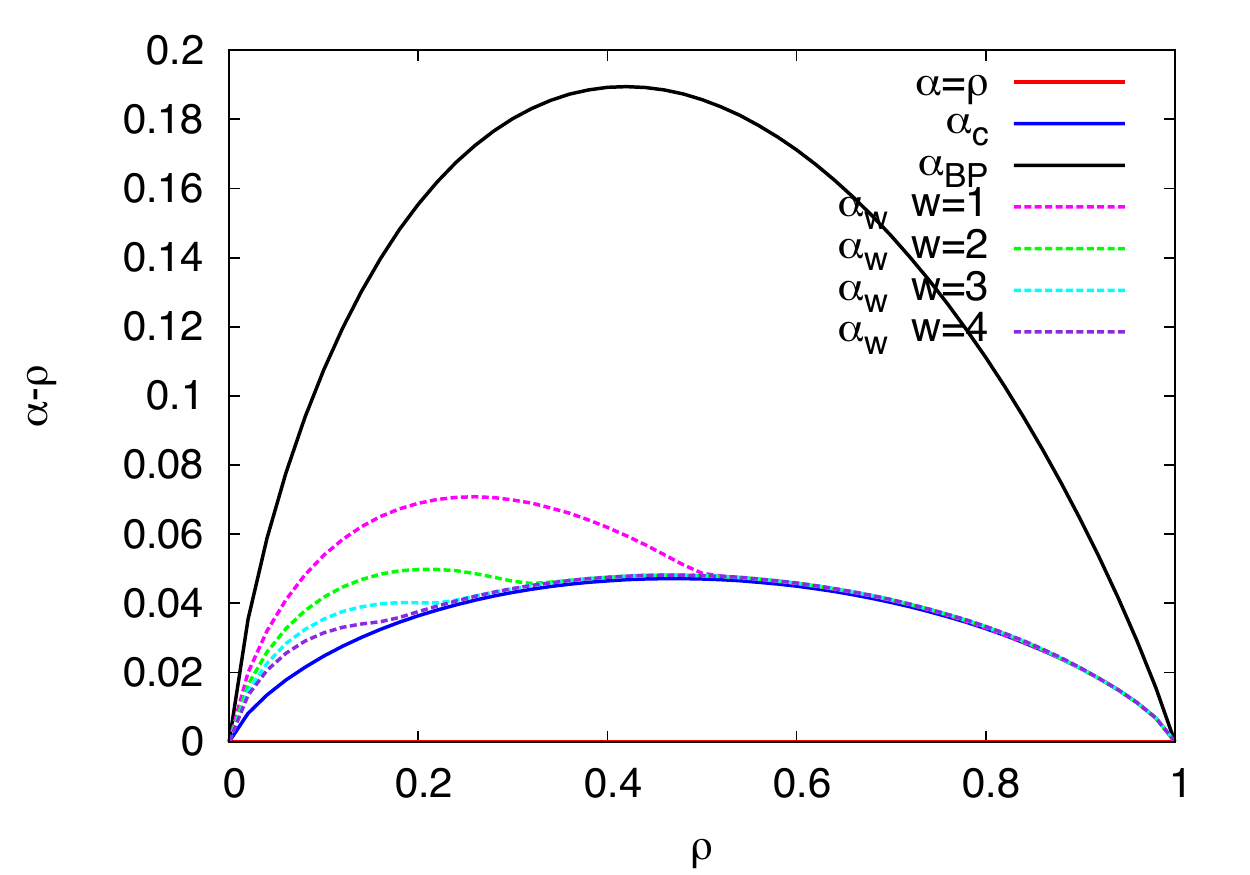}
\caption{A phase diagram for the coupled system with $\D=10^{-12}$,
  $L=240$. The black solid line (uppermost) and the blue solid line (lowermost) mark,
  respectively, the BP threshold and the reconstructability limits 
  of the single system. The dashed lines represent 
the reconstruction thresholds of the coupled system for different
values of the coupling range. Increasing the coupling range, the
threshold of the coupled system approaches the critical line of the
single system maintaining a smaller and smaller bump at small $\r$.
Note that the Donoho-Tanner phase transition \cite{Donoho05072005} well
known in compressed sensing literature would be considerably above the
$\alpha_{\rm BP}$ line.}
\label{varw}
\end{figure}

In Fig.~\ref{varw} we draw the $\alpha_w$ transition line (above which
reconstruction with spatially coupled matrix is achieved) for
measurement noise $\D=10^{-12}$ and different values of the interaction range from $w=1$ to $w=4$. We notice that for low values of $\r$, the transition line of the coupled system $\a_w$ is 
well separated from the critical line (reconstructability limit)
$\alpha_c$ of the single system. Whereas for larger values of $\rho$
the threshold $\alpha_w$ is extremely close to $\alpha_c$. The values of $\alpha_c$ used
here are computed with the equations resulting from the replica method calculation
from~\cite{KrzakalaMezard12}. 

We are not aware of previous work where this observation of two
separate regimes (large versus small gap from optimality at small
versus large values of sparsity $\rho$) existing for small $w$ would
have been made. Interestingly, careful examination of the data presented in Fig.~2 of
\cite{KrzakalaPRX2012} leads to a conclusion that also in that case
(also $w=1$) this effect is visible. 
With growing $w$, instead, the transition line tends to lean over the critical line of the single system, until 
they will be identical for $w\gg 1$.

\begin{figure}
\centering
\includegraphics[scale = 0.7]{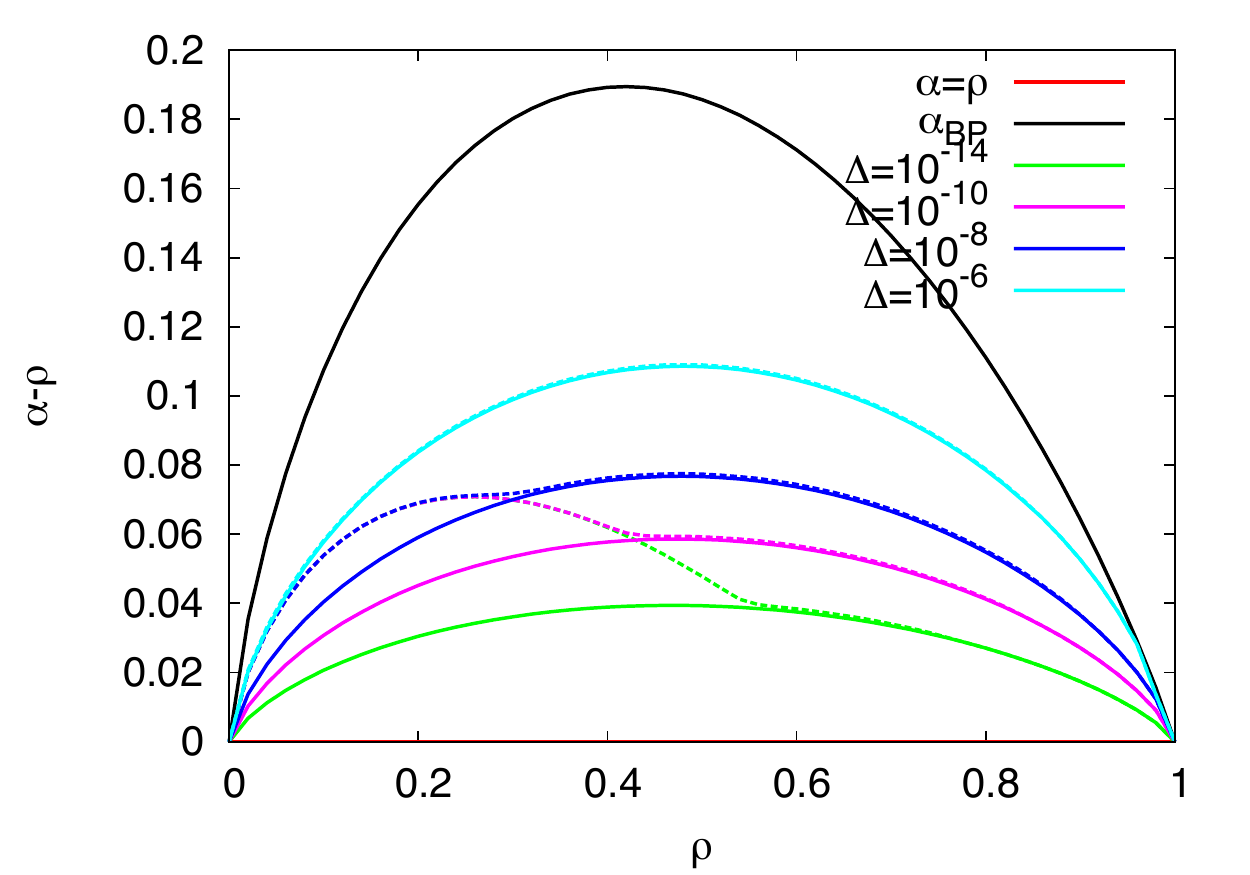}
\caption{A phase diagram for the coupled system with $w=1$,
  $L=240$. The black solid line (uppermost) marks the BP threshold of the
  single 
  system (interestingly this line does not change visibly for the considered values od $\Delta$). The solid and dashed lines in other colors represent,
  respectively, the critical line of the single system and the threshold
  for the coupled system for different noise levels.}
\label{vardelta}
\end{figure}

In Fig.~\ref{vardelta} we give the same diagram as in Fig.~\ref{varw}, but this time at fixed $w=1$ and for different 
values of the noise variance~$\D$. In this case the line
$\alpha_w$ seems to have low-$\rho$ part that is roughly
$\Delta$-independent (similarly to the $\alpha_{\mathrm{BP}}$ threshold), and a high-$\rho$ part that is always very close
to the ($\Delta$-dependent)
reconstructability limit $\alpha_c$. 

For the coupled system, we can write the potential (negative {\it free energy} in physics) as a function 
of the error profile $\lbrace E_p \rbrace$~\cite{KrzakalaMezard12},
the fixed points of the state evolution being given by the stationary points of the potential. 
As explained in \cite{Urbankechains} for a simpler system, namely the Curie-Weiss model, we can also obtain 
the free energy $\Phi (E)$ as a function of the mean-squared error by
extremizing the ``profile'' free energy subject to the constraint
$E=\sum_r E_r/L$. In \cite{Urbankechains} the authors also show that the potential
function $\Phi(E)$ of the spatially coupled system in the ordered
limit $L\to \infty$, 
$w \to \infty$ becomes the convex envelope of the free energy of
the single block. Exactly at the critical ratio $\a_b=\a_c$ the two
maxima of the potential function corresponding to optimal reconstruction and failed reconstruction respectively, are connected by a {\it plateau}. This means that, 
for any $\a_b$ infinitesimally larger than $\a_c$, there is only one
maximum of the potential, and it gives the MSE of the 
optimal reconstruction. This is the explanation (in terms of the potential) of the fact that the threshold line for the coupled system tends 
to the critical line of the single system when $w\to \infty$.

The way in which the potential behaves at finite $w$ (\ie the way in which it converges to the convex envelope)
 determines the structure of the threshold lines in Figs.~\ref{varw} and \ref{vardelta}. When the system is coupled 
 with growing $w$, the potential of the single system tends to be flattened in the region in between the two 
 fixed points and it is ``dressed'' with oscillations whose frequency increases with $L$ and whose amplitude 
 decreases with both $L$ and $w$. In our case we see that the convergence is faster for larger $\r$ and is much slower 
 for small $\r$. When the finite-$w$ threshold line is very close to the critical line of the single system 
 the scenario is very similar to the one presented in \cite{Urbankechains,CaltaCurie} for the Curie-Weiss model. 
 On the other hand, when the separation between $\a_w$ and $\a_c$ is
 large, as it is for $w=1$, low $\rho$ and small $\D$, convergence of
 $\alpha_w\to \alpha_c$ becomes much slower in $w$. The fundamental
 reason behind this remains to be unveiled in future work.

\section{The seeding condition} 

What we presented in the previous Section is valid for very large seed $w_s$ and 
and very large seeding ratio $\a_s$. The seed has to be present in order to trigger the
propagation of the reconstruction wave. However, given a fixed bulk-measurement-ratio $\a_b\in [\a_w,\a_{BP}]$, 
the effective measurement ratio (average over the whole system)
$\a_{\mathrm{eff}}$ will be slightly bigger than $\a_b$ and therefore
for finite $L$ we will not be able to access values of $\a_{\mathrm{eff}}$ exactly down to $\a_w$. 

\begin{figure}
\centering
\includegraphics[scale = 0.7]{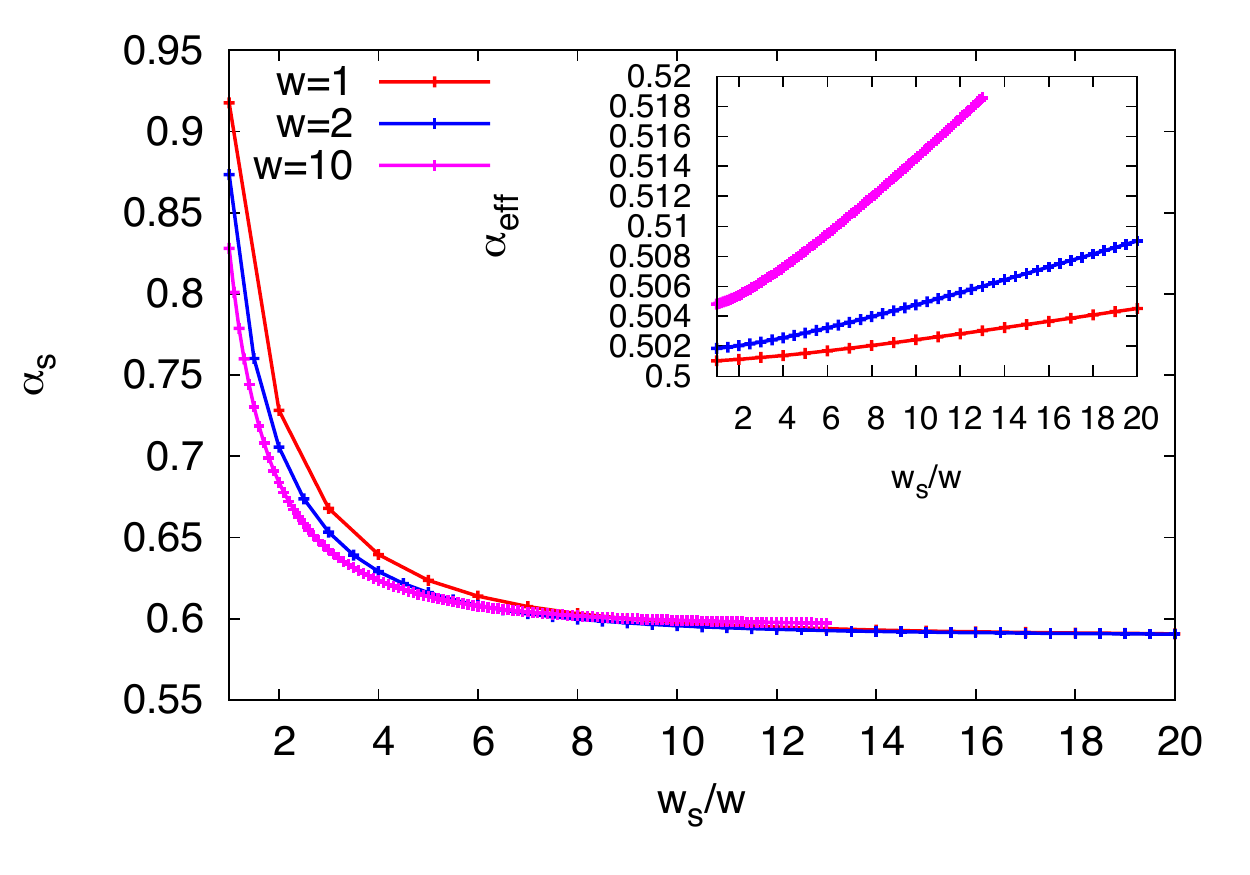}
\caption{[Main frame] The wave propagation/non-propagation diagram as a
  function of the seed size $w_s$ and seed strength $\alpha_s$ 
for $\r=0.4$, $\D=10^{-12}$, $\a_b=0.5$ and $L=400$ for three different values of the interaction range $w=1,2,10$. Below the lines the reconstruction wave does not propagate while above it does. [Inset] The effective measurement ratio along the transition line for the same values of the interaction range.}
\label{ph_diag_flat}
\end{figure}

As an example, in Fig.~\ref{ph_diag_flat} we show the propagation/non-propagation phase diagram as a function 
of the seed size and strength for $w=1,2,10$, $\r=0.4$, $\D=10^{-12}$,
$\a_b=0.5$ and $L=400$. Just as it happens in the case of the spatially-coupled Curie-Weiss model \cite{CaltaCurie}, the transition line approaches a smooth limiting curve when the interaction range $w$ grows. In the inset we show the effective measurement ratio $\a_{\mathrm{eff}}$ 
along the transition line, \ie the best we can do given the bulk value~$\a_b$, the interaction range $w$ and the fixed number of blocks. Again, as in the Curie-Weiss model \cite{CaltaCurie}, the best choice (the one that minimizes $\a_{\mathrm{eff}}$) is $w=1$. 
Notice that for $\r=0.4$, $\D=10^{-12}$ and $w=1$, we have $\a_w=0.4619$, so the value of $\a_b$ we have chosen 
is above the threshold line for unitary interaction range. In other cases, as for example $\r=0.2$, $\D=10^{-12}$ and $\a_b=0.26$, 
the wave does not propagate with $w=1$ for any choice of the seed as we can see from the diagram of Fig.~\ref{varw} and 
the best choice becomes therefore $w=2$.

\section{The role of the interaction shape}
Until now we have considered the case of a constant interaction $g(x)=1/2$ on the interval $[-1,1]$. 
On the other hand, in our model any properly normalized function is allowed. A~first simple generalization is an interaction of the form 
\beq
g(x)=\frac{1}{2}+Ax
\eeq
with $A \in [-1/2,1/2]$. 
\begin{figure}
\centering
\includegraphics[scale = 0.7]{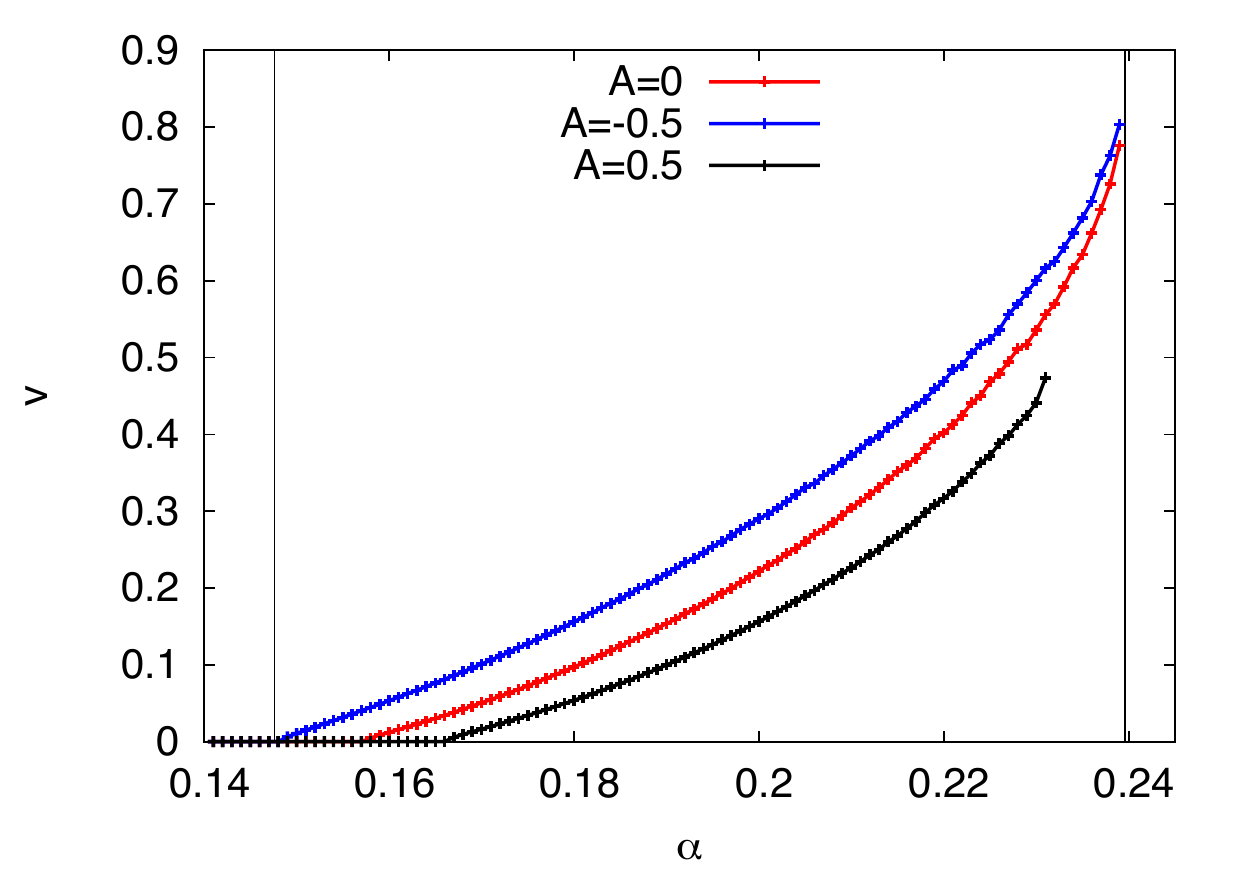}
\caption{The propagation speed in the interval $\a_b \in
  [\a_c,\a_{BP}]$ for $\r=0.12$ and $\D=10^{-12}$, $w=3$, $L=400$. We report 
the speed for three different angular coefficients of the interaction function $g(x)$, namely $A=-0.5,0,0.5$. The two 
vertical lines mark $\a_c$ (left) and $\a_{\mathrm{BP}}$ (right). Note
that the transition $\alpha_w$ (when the speed gets positive) depends
considerably on $A$.}
\label{speed}
\end{figure}

The value of the constant $A$ influences three different quantities, namely: 
\begin{itemize}
\item the position of the finite-$w$ transition $\a_w$ 
\item the speed $v$ of the propagating wave defined as the number of reconstructed blocks per iteration
\item the necessary size and strength of the seed
\end{itemize}
All the three characteristics are in favor of a negative $A$ coefficient, meaning that each block interacts more strongly backwards than forwards. In this case, in fact, the finite-$w$ transition moves closer to $\a_c$, the propagation speed is increased and the necessary seeding is weaker. Viceversa, when the angular coefficient is positive, 
the propagation speed is lower and the seeding conditions must be stronger. 

Note that in the original work on spatial coupling in compressed
sensing \cite{KrzakalaPRX2012,KrzakalaMezard12} the interaction was chosen strongly asymmetric, our
present study confirms this choice as the more efficient one. 

\begin{figure}
\centering
\includegraphics[scale = 0.7]{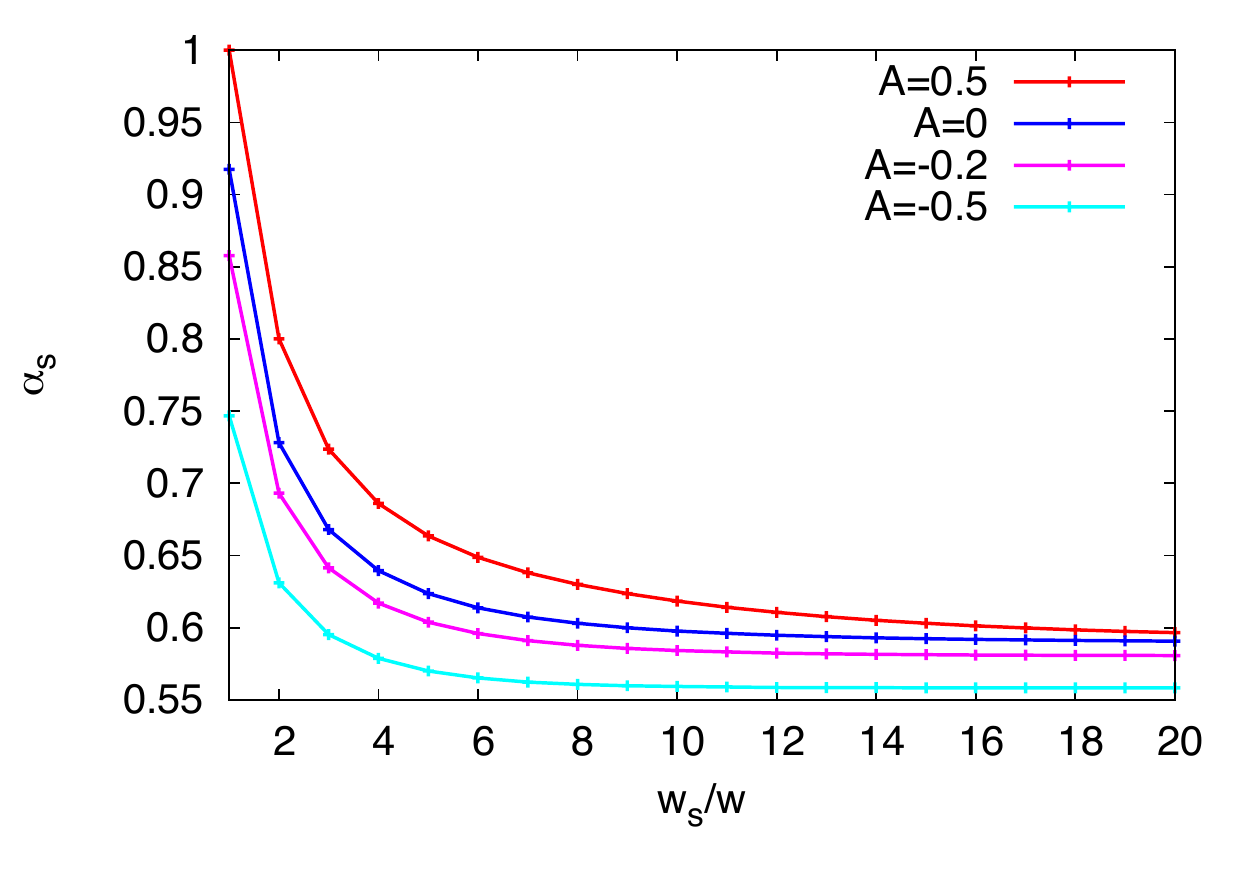}
\caption{The propagation phase diagram as a function of the seed features for different values of the angular 
coefficient in the interaction function $g(x)$, with $\r=0.4$,
$\D=10^{-12}$, $\a_b=0.5$, $w=1$ and $L=400$.}
\label{seed_tilt}
\end{figure}

In Fig.~\ref{speed} we show the propagation speed in the interval $\a_b \in [\a_c,\a_{BP}]$ for three different values of $A$, while in Fig.~\ref{seed_tilt} we report the seeding diagrams for several values of the angular coefficient. 

\section{Conclusion}
In this work we have studied the design of spatially coupled measurement matrices for compressed sensing. 
In particular we have chosen the structure defined in
Fig.~\ref{matrix}, and evaluated its efficiency as a function 
of the interaction range~$w$, the interaction shape $g(x)$, and the
size and strength of the seed $w_s$ and $\alpha_s$. 
We find that, for sufficiently large seed, there is a
threshold line $\alpha_w$ for the coupled system below which 
reconstruction fails. This line lies in between the lowest possible
subsampling rate at which Bayes-optimal inference succeeds $\alpha_c$
and the $\alpha_{\rm BP}$ threshold of the single system.  
The line $\alpha_w$ is considerably separated from $\alpha_c$ when the
interaction range $w$ and density $\rho$ are both sufficiently small. 
We also studied the optimality of the seeding condition, concluding
that for small termination cost small range $w$ is better.  
Finally we analyzed the speed of the reconstruction as a function of
the interaction shape, given by the function $g(x)$ and concluded that
a strong asymmetry is advantageous. 

\section*{Acknowledgment}
This work has been supported in part by the ERC under the European
Union’s 7th Framework Programme Grant Agreement 307087-SPARCS, and by
the project TASC of the Labex ‘‘PALM.’’

\bibliographystyle{IEEEtran}
\bibliography{refs}

\begin{thebibliography}{10}
\providecommand{\url}[1]{#1}
\csname url@samestyle\endcsname
\providecommand{\newblock}{\relax}
\providecommand{\bibinfo}[2]{#2}
\providecommand{\BIBentrySTDinterwordspacing}{\spaceskip=0pt\relax}
\providecommand{\BIBentryALTinterwordstretchfactor}{4}
\providecommand{\BIBentryALTinterwordspacing}{\spaceskip=\fontdimen2\font plus
\BIBentryALTinterwordstretchfactor\fontdimen3\font minus
  \fontdimen4\font\relax}
\providecommand{\BIBforeignlanguage}[2]{{%
\expandafter\ifx\csname l@#1\endcsname\relax
\typeout{** WARNING: IEEEtran.bst: No hyphenation pattern has been}%
\typeout{** loaded for the language `#1'. Using the pattern for}%
\typeout{** the default language instead.}%
\else
\language=\csname l@#1\endcsname
\fi
#2}}
\providecommand{\BIBdecl}{\relax}
\BIBdecl

\bibitem{FelstromZigangirov99}
A.~Jimenez~Felstrom and K.~Zigangirov, ``Time-varying periodic convolutional
  codes with low-density parity-check matrix,'' \emph{IEEE Transactions on
  Information Theory}, vol.~45, no.~6, pp. 2181 --2191, 1999.

\bibitem{lentmaier2005terminated}
M.~Lentmaier, A.~Sridharan, K.~S. Zigangirov, and D.~Costello, ``Terminated
  ldpc convolutional codes with thresholds close to capacity,'' in
  \emph{Information Theory, 2005. ISIT 2005. Proceedings. International
  Symposium on}.\hskip 1em plus 0.5em minus 0.4em\relax IEEE, 2005, pp.
  1372--1376.

\bibitem{lentmaier2010iterative}
M.~Lentmaier, A.~Sridharan, D.~J. Costello, and K.~S. Zigangirov, ``Iterative
  decoding threshold analysis for ldpc convolutional codes,'' \emph{IEEE
  Transactions on Information Theory}, vol.~56, no.~10, pp. 5274--5289, 2010.

\bibitem{KudekarRichardson10}
S.~Kudekar, T.~J. Richardson, and R.~L. Urbanke, ``Threshold saturation via
  spatial coupling: Why convolutional ldpc ensembles perform so well over the
  bec,'' \emph{IEEE Transactions on Information Theory}, vol.~57, no.~2, pp.
  803--834, 2011.

\bibitem{KudekarRichardson12}
S.~Kudekar, T.~Richardson, and R.~Urbanke, ``Spatially coupled ensembles
  universally achieve capacity under belief propagation,'' in \emph{Information
  Theory Proceedings (ISIT), 2012 IEEE International Symposium on}.\hskip 1em
  plus 0.5em minus 0.4em\relax IEEE, 2012, pp. 453--457.

\bibitem{Donoho:06}
D.~L. Donoho, ``{Compressed sensing},'' \emph{IEEE Trans. Inform. Theory},
  vol.~52, p. 1289, 2006.

\bibitem{Candes:2008}
E.~J. Cand{\`e}s and M.~B. Wakin, ``An introduction to compressive sampling,''
  \emph{IEEE Signal Processing Magazine}, vol.~25, no.~2, pp. 21--30, Mar.
  2008.

\bibitem{KudekarPfister10}
S.~Kudekar and H.~Pfister, ``The effect of spatial coupling on compressive
  sensing,'' in \emph{Communication, Control, and Computing (Allerton)}, 2010,
  pp. 347--353.

\bibitem{KrzakalaPRX2012}
F.~Krzakala, M.~M{\'e}zard, F.~Sausset, Y.~Sun, and L.~Zdeborov{\'a},
  ``Statistical physics-based reconstruction in compressed sensing,''
  \emph{Phys. Rev. X}, vol.~2, p. 021005, 2012.

\bibitem{KrzakalaMezard12}
F.~Krzakala, M.~M\'ezard, F.~Sausset, Y.~Sun, and L.~Zdeborov\'a,
  ``Probabilistic reconstruction in compressed sensing: Algorithms, phase
  diagrams, and threshold achieving matrices,'' \emph{J. Stat. Mech.}, p.
  P08009, 2012.

\bibitem{DonohoJavanmard11}
D.~L. Donoho, A.~Javanmard, and A.~Montanari, ``Information-theoretically
  optimal compressed sensing via spatial coupling and approximate message
  passing,'' in \emph{Proc. of the IEEE Int. Symposium on Information Theory
  (ISIT)}, 2012.

\bibitem{YedlaJian12}
A.~Yedla, Y.-Y. Jian, P.~S. Nguyen, and H.~D. Pfister, ``A simple proof of
  threshold saturation for coupled scalar recursions,'' in \emph{7th
  International Symposium on Turbo Codes and Iterative Information Processing
  (ISTC)}.\hskip 1em plus 0.5em minus 0.4em\relax IEEE, 2012, pp. 51--55.

\bibitem{DonohoMaleki09}
D.~L. Donoho, A.~Maleki, and A.~Montanari, ``Message-passing algorithms for
  compressed sensing,'' \emph{Proc. Natl. Acad. Sci.}, vol. 106, no.~45, pp.
  18\,914--18\,919, 2009.

\bibitem{Rangan10b}
S.~Rangan, ``Generalized approximate message passing for estimation with random
  linear mixing,'' in \emph{IEEE International Symposium on Information Theory
  Proceedings (ISIT)}, 2011, pp. 2168 --2172.

\bibitem{bayati2012universality}
M.~Bayati, M.~Lelarge, and A.~Montanari, ``Universality in polytope phase
  transitions and message passing algorithms,'' \emph{arXiv preprint
  arXiv:1207.7321}, 2012.

\bibitem{kudekar2012wave}
S.~Kudekar, T.~Richardson, and R.~Urbanke, ``Wave-like solutions of general
  one-dimensional spatially coupled systems,'' \emph{arXiv preprint
  arXiv:1208.5273}, 2012.

\bibitem{Donoho05072005}
D.~L. Donoho and J.~Tanner, ``Sparse nonnegative solution of underdetermined
  linear equations by linear programming,'' \emph{Proceedings of the National
  Academy of Sciences of the United States of America}, vol. 102, no.~27, pp.
  9446--9451, 2005.

\bibitem{Urbankechains}
S.~H. Hassani, N.~Macris, and R.~Urbanke, ``Chains of mean-field models,''
  \emph{Journal of Statistical Mechanics: Theory and Experiment}, vol. 2012,
  no.~02, p. P02011, 2012.

\bibitem{CaltaCurie}
F.~Caltagirone, S.~Franz, R.~Morris, and L.~Zdeborov\'a, ``Dynamics and
  termination cost of spatially coupled mean-field models,''
  \emph{arXiv:1310.2121v1}, 2013.

\end{thebibliography}

\end{document}